\pdfoutput = 1
\documentclass[%
 reprint,
 prd,
 amsmath,amssymb,
 aps,
 superscriptaddress,
 floatfix
]{revtex4-1}

\usepackage{dcolumn}
\usepackage{bm}
\usepackage{ulem}
\usepackage{multirow}
\usepackage{placeins}
\usepackage{algorithmic}
\usepackage{color}
\usepackage{comment}
\usepackage{hyperref}

\usepackage{array}
\newcolumntype{P}[1]{>{\centering\arraybackslash}p{#1}}
\begin{document}

\title{Quasinormal modes of anyons}
\author{Vishnulal C}\thanks{lal23@iisertvm.ac.in}
\affiliation{School of Physics, Indian Institute of Science Education and Research Thiruvananthapuram, Maruthamala PO, Vithura, Thiruvananthapuram 695551, Kerala, India}
\author{Saurya Das}\thanks{saurya.das@uleth.ca}
\affiliation{Theoretical Physics Group and Quantum Alberta, Department of Physics and Astronomy, University of Lethbridge 4401 University Drive, Lethbridge, Alberta T1K 3M4, Canada}
\author{Soumen Basak}\thanks{sbasak@iisertvm.ac.in}
\affiliation{School of Physics, Indian Institute of Science Education and Research Thiruvananthapuram, Maruthamala PO, Vithura, Thiruvananthapuram 695551, Kerala, India}
\date{\today}

\begin{abstract}
We derive the quasinormal modes of anyons for $(2+1)$-dimensional
Bañados, Teitelboim, and Zanelli (BTZ) and analogue black holes
and discuss potential experiments to measure these modes. 
\end{abstract}

\maketitle

\setlength{\columnsep}{0.5cm}

\section{Introduction}
Black holes, predicted by the general theory of relativity, are often considered to be laboratories for testing theories of classical and quantum gravity. Two outstanding unresolved issues in these theories are the resolution of black hole singularities and the information loss problem. Although there have been a number of interesting proposals towards their resolution \cite{https://doi.org/10.48550/arxiv.hep-th/9508151,Mathur:2005zp,Almheiri:2013hfa,doi:10.1142/S0218271821500693}, one still lacks a complete resolution. To better understand the issues, it is important to study a variety of phenomena that black holes demonstrate and their implications. Quasinormal modes (QNM) of black holes is one such important phenomenon to study in this context \cite{PhysRevD.93.044048,PhysRevD.105.044015,PhysRevLett.123.111102}.

Quasinormal modes (QNM) are generated when a black hole is perturbed by a scalar, vector, or tensor disturbance, resulting in a decaying ringing signal \cite{Chandrasekhar:1975zza}. The real and imaginary components of these modes denote the ringing frequency and the reciprocal of the characteristic decay time, respectively. These characteristics of QNMs can provide valuable insights into black holes and their interiors. For instance, they can be employed in the study of gravitational waves produced by binary black hole mergers \cite{2008qnm}. The outcomes derived from these studies constitute some of the most critical assessments of general relativity and the properties of black holes.

Perhaps the most important fact is that the frequency of Quasinormal Modes (QNM) depends only on the black hole parameters, and therefore carries rich information about the black holes themselves. This, coupled with the fact that a black hole can be fully characterized by a few parameters such as mass ($M$), charge ($Q$), and angular momentum ($J$), QNM offers a 'window' into what completely characterizes a black hole, whether it is astrophysical or an analogue black hole produced in the laboratory.

Thanks to an intriguing connection between quasinormal modes and quantum gravity, there has been a surge of interest in the QNM of scalar, electromagnetic, and other fields from a large class of black holes \cite{Panotopoulos:2018can,churilova2020quasinormal,PhysRevD.101.124012,2004}. However, to the best of our knowledge, QNM of \textit{anyonic} excitations of a black hole has not been studied so far. This is an important avenue to explore because, on the one hand, anyons following fractional statistics can be produced and exist in $(2+1)$-dimensional spacetime; in fact, there has been an active interest in their search recently \cite{doi:10.1126/science.aaz5601}, while on the other hand, analogue black holes have also been produced in $(2+1)$-dimensions and provide a natural space for testing various properties of anyons \cite{article32,article40,article22}.

We are thus motivated by two things. First, testing the existence of anyons and studying their properties, and second, studying properties of $(2+1)$-dimensional black holes and their analog counterparts, as probed by anyons. Anyons are most likely to be found in two-dimensional condensed matter systems. This is also relevant given the importance and recent interest in anyons and their applications, for example, in the field of topological quantum computation. It provides a setup for constructing a new quantum algorithm inspired by the evolution of topological excitations \cite{RevModPhys.80.1083,Field_2018}. Therefore, studying all the aspects of anyons, including their behavior in a host of systems (such as studied here), is expected to shed light on the particles themselves. This is also an appropriate juncture for doing these investigations since there is already experimental evidence in favor of anyons and their braiding statistics in the $\nu=\frac{1}{3}$ fractional quantum Hall state of anyons \cite{Nakamura_2020}. The experimental proposal that we discuss in our paper (and a similar proposal in our earlier paper \cite{PhysRevD.104.104011}) adds to the above effort of gathering evidence for anyons.

Our second motivation is to study the properties of quantum black holes, or more precisely, quantum fields in the background of black hole horizons, in relation to phenomena such as Hawking radiation and quasinormal modes. This research is useful in understanding and eventually resolving the black hole information loss problem and will shed important light on the development of quantum gravity itself. Since such quantum properties are almost impossible to study in the astrophysical scenario with current technologies, our only hope is to study their analogues, as some of the results obtained there should continue to hold for real black holes, such as the thermal spectrum for Hawking radiation. An analogy between anyon Van der Waals fluid and AdS anyon Van der Waals black hole was also studied in \cite{AGHAEIABCHOUYEH2018240}, and anyon-like excitations in the context of Bañados, Teitelboim, and Zanelli (BTZ) black holes in \cite{Luo:2017ksc}. In this respect, we noted in our earlier paper \cite{PhysRevD.104.104011} that the behaviors of bosons, fermions, and anyons in Hawking radiation are quite different, and each provides important and complementary information about the black hole. For quasinormal modes too, it is known that bosonic and fermionic results differ \cite{PhysRevD.97.084034,https://doi.org/10.48550/arxiv.2210.14973}. Therefore, it is important to look at other excitations in similar backgrounds, as we have done in this work.

The remainder of this paper is organized as follows. In Section \ref{anyons}, we provide a brief introduction to anyons, focusing on properties relevant to our work. This is followed by a concise review in Section \ref{QNMscalar} of the Quasinormal Modes (QNM) for a scalar field in a BTZ background. We choose BTZ as a representative and well-understood $(2+1)$-dimensional black hole, which has also been the subject of QNM studies for non-anyonic fields \cite{2017, Panotopoulos:2018can}. Sections \ref{qnansection} and \ref{experimentalset-up} are dedicated to deriving the QNM frequencies of anyons for BTZ black holes and adapting our findings to analogue black holes that can be created in laboratory settings. We also explore methods for measuring the QNM. In conclusion, we summarize our work and outline open problems in Section \ref{conclusionsection}.

\section{Anyons}
\label{anyons}

In three-dimensional space, all particles fall into two categories: bosons and fermions, distinguished by their spins, which are either integral or half-integral values in units of $\hbar$. These particles follow distinct statistics: Bose-Einstein statistics for bosons and Fermi-Dirac statistics for fermions. Their wavefunctions exhibit different symmetries, with bosonic wavefunctions being symmetric, and fermionic wavefunctions being anti-symmetric when particles are exchanged. However, in the context of $(2+1)$-dimensional spacetime, the restrictions on particle statistics become more flexible \cite{Sen:1993qc}. This flexibility can be readily demonstrated by examining the behavior of the wavefunction for two identical particles in $(2+1)$-dimensional spacetime. Let $\psi(r)$ represent the wave function of a system comprising these two particles, subject to the condition that $\psi(r) \neq 0$ for $r > a$, where $a$ corresponds to the 'hard-core condition.' Here, $\vec{r}_1$ and $\vec{r}_2$ denote the position vectors of the two particles, and $\vec{r} \equiv \vec{r}_1 - \vec{r}_2$ signifies the relative position vector. Consequently, the configuration space for these particles is the two-dimensional $(x, y)$-plane with a disc of radius $a$ removed.

Defining a complex coordinate $z = x + iy$ and performing a transformation $z\rightarrow z\,e^{2\pi i}$, which effectively returns the particle to its initial position, the wave function should remain invariant, albeit with a phase factor. In other words, we have

\begin{equation}
   \psi(z e^{2\pi i}, z^{*} e^{-2\pi i}) = e^{2\pi i \alpha} \psi(z, z^{*}),
    \label{Eq1}
\end{equation}

where $\alpha$ is a real parameter.

Similarly, it is possible to interchange the positions of two particles (i.e., transform $z \rightarrow z e^{\pi i}$), leading to the following relationship:

\begin{equation}
    \psi(z e^{\pi i}, z^{*} e^{-\pi i}) = e^{\pi i \alpha} \psi(z, z^{*})~.
    \label{Eq2}
\end{equation}
In this case, too, $\alpha$ is a real parameter. The value of $\alpha$ takes on the specific values of $0$ and $1$ for bosons and fermions, respectively. However, in the context of $(2+1)$-dimensions, $\alpha$ can assume any real value between $0$ and $1$. These intermediate values of $\alpha$ correspond to an entirely new category of particles with fractional statistics and are referred to as "anyons."
\section{Quasinormal modes of the massless scalar field from BTZ black hole}
\label{QNMscalar}

The action for a massless scalar field $\Phi$ with a non-zero coupling $\epsilon$ to the Ricci scalar $R_{3}$ in $(2+1)$ dimensions can be written as:

\begin{equation}
    S = \frac{1}{2}\int d^{3}x \sqrt{-g}\left[\partial_{\mu}\Phi\partial^{\mu}\Phi+\epsilon\,R_{3}\Phi^{2} \right].
\end{equation}

The variation of this action with respect to the field $\Phi$ gives the corresponding equation of motion:

\begin{equation}
    \frac{1}{\sqrt{-g}}\partial_{\mu}\left(\sqrt{-g}g^{\mu \nu}\partial_{\nu}\right)\Phi=\epsilon R_{3}\Phi.
    \label{eqmotionscalar}
\end{equation}

The BTZ black hole \cite{PhysRevLett.69.1849} is a solution of $(2+1)$-dimensional topological gravity with a negative cosmological constant. Its metric can be written as:

\begin{equation}
    ds^{2}=-f(r)dt^{2}+\frac{dr^{2}}{f(r)}+r^{2}d\theta^{2},
    \label{btzmetric}
\end{equation}

In this work, we consider a non-rotating BTZ black hole for which the function $f(r)$ depends on the mass $M$ of the black hole and cosmological constant, $\Lambda=-1/l^{2}$, namely:

\begin{equation}
    f(r)\,=\,-M\,+\,\frac{r^{2}}{l^{2}} \qquad R_{3}=-\frac{6}{l^{2}} = 6\Lambda.
    \label{eqf}
\end{equation}

Given the symmetry of the spacetime, we make the following ansatz for the scalar field $\Phi$:

\begin{equation}
    \Phi(r\,,t\,,\theta)\,=\,R(r)\,e^{-i\,\omega\,t}\,e^{i\,m\,\theta},
    \label{ansatz}
\end{equation}

where $\omega$ is the frequency and $m$ is the angular momentum quantum number. The modes with $\omega_{I} > 0$ represent unstable modes while modes with $\omega_{I} < 0$ represent stable modes.

Substituting the above ansatz into the equation of motion, we obtain the following equation satisfied by the radial part $R(r)$:

\begin{equation}
    R^{''}+\left(\frac{1}{r}+\frac{f^{'}}{f}\right)R^{'}+\left(\frac{\omega^{2}}{f^{2}}-\frac{m^{2}}{r^{2}f}+\frac{6\epsilon}{fl^{2}}\right)R=0.
    \label{Radialpartscalar}
\end{equation}

We can recast the above equation in the form of a Schrödinger-like equation and thereby separate out the effective potential that the scalar field feels. To this end, we first introduce a variable as follows:

\begin{equation}
    R\,=\,\frac{\chi}{\sqrt{r}}.
\end{equation}
Next, we introduce the tortoise coordinates $r_{*}$, which can be expressed as:
\begin{equation}
    r_{*}\,=\,\frac{l^{2}}{2r_{H}}\ln{\frac{r-r_{H}}{r+r_{H}}},
\end{equation}
where $r_{H}\,=\,l\sqrt{M}$ is the horizon radius of the BTZ black hole. With this, Eq.(\ref{Radialpartscalar}) takes the form of a Schrödinger-like equation in $r_{*}$ coordinates:
\begin{equation}
    \frac{d^{2}\chi}{dr_{*}^{2}}\,+\,\left(\omega^{2}-V(r)\right)\chi\,=\,0,
    \label{SE3}
\end{equation}
where the effective potential $V(r)$ is given by:
\begin{equation}
    V(r)\,=\,f(r)\left(-\frac{6\epsilon}{l^{2}}+\frac{m^{2}}{r^{2}}+\frac{f^{'}(r)}{2r}-\frac{f}{4r^{2}}\right).
    \label{pot1}
\end{equation}
For the solutions of Eq.(\ref{SE3}), we impose boundary conditions at the horizon and infinity \cite{PhysRevD.63.124015,PhysRevD.1.2870}. At $r=r_{H}$, we have a purely in-going wave $\sim e^{i\omega r_{*}}$, which represents a black hole, and at $r=\infty$, we have a purely out-going wave $\sim e^{-i\omega r_{*}}$, to exclude any incoming radiation.

Furthermore, Eq.(\ref{Radialpartscalar}) can be converted to a Gauss’ hypergeometric equation by the introduction of suitable parameters. The solution of the hypergeometric equation, together with the aforementioned boundary conditions, gives us the exact spectrum
of QNM by solving Eq.(\ref{Radialpartscalar})\cite{Panotopoulos:2018can}.
\begin{equation}
    \omega_{n}=\omega_{R}\,+\,i\,\omega_{I}~,
\end{equation}
where $\omega_{R}$ and $\omega_{I}$ respectively are the real and imaginary parts of the quasinormal frequencies,
\begin{eqnarray}
    \omega_{R}&=&\left| \,\frac{\mid m \mid}{l}\,-\,\frac{\sqrt{M}}{l}\,\sqrt{6\,\epsilon\,-\,1}\,\right|\label{realpt}\\
    \omega_{I}&=&-\frac{2\sqrt{M}}{l}\,\left(n\,+\frac{1}{2}\right)~.
    \label{imagpt}
\end{eqnarray}
Negative values of $\omega_{I}$ clearly indicate that the corresponding QNM are stable modes. It should be noted that in the above derivation, an assumption of $\epsilon > 1/6$ was required to be made, consistency with which is also reflected from the square root term in  Eq.(\ref{realpt}). Similar solutions 
can be obtained when $\epsilon \leq 1/6$
\cite{PhysRevD.64.064024}.
\section{Quasinormal modes of anyons from BTZ black hole}
\label{qnansection}
In this section, we will extend the derivation of Quasinormal Modes (QNM) associated with a BTZ black hole for a massless scalar field to the case of an anyonic field. Consider an anyonic field with a nonzero coupling $\epsilon$ to the Ricci scalar described by the following action:
\begin{equation}
    S = \frac{1}{2}\int d^{3}x\sqrt{-g} \left[L + \epsilon R_{3}\Phi^{2}\right]~,
    \label{act1}
\end{equation}
where $L$ is the Lagrangian for the anyonic field $\Phi$ and $R_{3}$ is the constant Ricci scalar of the BTZ background. We will consider a non-rotating ($J=0$) BTZ black hole background given by the metric in Eqs. (\ref{btzmetric}) and (\ref{eqf}). It may be noted that the Lagrangian represents an abelian Higgs model with a Chern-Simons (CS) term, whose classical solutions will give rise to anyonic excitations and is given by \cite{Rao1992AnAP}:
\begin{multline}
    L = -\frac{1}{4}F_{\mu\nu}F^{\mu\nu} + \frac{1}{2}(\partial_{\mu}-iqA_{\mu})\Phi^{*}(\partial^{\mu}+iqA^{\mu})\Phi \\
    - c_{4}\left(\Phi \Phi^{*}-\frac{c_{2}}{2c_{4}}\right)^{2} + \frac{\mu}{4}\epsilon_{\mu\nu\alpha}F^{\mu\nu}A^{\alpha}~,
    \label{lag1}
\end{multline}
where $F_{\mu\nu}$ is the electromagnetic field tensor, $A_{\mu}$ is the gauge field, and $c_{2}$ and $c_{4}$ are real constants. The last term in the Lagrangian is the added CS term, and the total Lagrangian is gauge-invariant.

The corresponding equation of motion for the anyonic field is as follows:
\begin{multline}
g^{\mu\nu}\nabla_{\mu}\nabla_{\nu}\Phi + 2iqg^{\mu \nu}A_{\mu}\partial_{\nu}\Phi \\
-\left(2c_2+q^2g^{\mu \nu} A_{\mu} A_{\nu}\right) \Phi + 4 c_4 \Phi^* \Phi^2 = \epsilon R_{3}\Phi~.
\label{eqann13}
\end{multline}

For our work, we will consider the following simple case:
\begin{eqnarray}
c_{2} = c_{4} = 0  \hspace{30pt} A_{\mu}=(0,0,r^{2})~,
\end{eqnarray}
and use the following ansatz for the anyonic field:
\begin{eqnarray}
    \Phi(r,t,\theta) = R(r)e^{-i\omega t}e^{im\theta}~,
    \label{ansatz102}
\end{eqnarray}
where $\omega$ and $\theta$ have the same meaning as in Section \ref{QNMscalar}.

The differential equation for the radial part of the anyonic field is as follows:
\begin{multline}
  R^{''} + \left(\frac{1}{r} + \frac{f^{'}}{f}\right)R^{'} \\
  + \left(\frac{\omega^{2}}{f^{2}} - \frac{m^{2}}{r^{2}f} + \frac{6\epsilon}{f l^{2}} - \frac{2qm}{f} - \frac{q^{2}r^{2}}{f}\right)R = 0 ~.
\end{multline}

We now introduce a dimensionless parameter $z$ related to $r$ as follows:
\begin{equation}
    z = 1 - \frac{r_{H}^{2}}{r^{2}}  \hspace{15pt} \Rightarrow \hspace{15pt}  r = \frac{r_{H}}{(1-z)^{\frac{1}{2}}}~,
\end{equation}
which converts the above differential equation to the following form:
\begin{multline}
    z(1-z)R_{zz} + (1-z)R_{z} \\
    + \left(\frac{A}{z} + \frac{B}{-1+z} - C + \frac{D}{1-z} + \frac{E}{(1-z)^{2}}\right)R = 0~,
\end{multline}
where
\begin{multline}
 A = \frac{l^{4}\omega^{2}}{4r_{H}^{2}}\hspace{15pt} B = -\frac{3\epsilon}{2} \hspace{15pt} C = \frac{l^{2}m^{2}}{4r_{H}^{2}} \\
  D = -\frac{q\, m\, l^{2}}{2} \hspace{15pt} E = -\frac{q^{2}l^{2}r_{H}^{2}}{4} ~.
\end{multline}

Next, we use the following ansatz: $R = z^{\alpha}(1-z)^{\beta}F$, with the approximation $q^{2}\sim0$, to convert it into the form of a Gauss’ hypergeometric equation:
\begin{multline}
    z(1-z)F_{zz}+[1+2\alpha-(1+2\alpha+2\beta)z]F_{z} \\
    + \left(\frac{\Bar{A}}{z} + \frac{\Bar{B}-D}{-1+z} - \Bar{C}\right)F = 0~,
\end{multline}
where
\begin{equation}
    \Bar{A} = A + \alpha^{2}  \hspace{15pt} \Bar{B} = B + \beta - \beta^{2} \hspace{15pt}
    \Bar{C} = C + (\alpha + \beta)^{2}~.
\end{equation}

We can choose the values of $\alpha$ and $\beta$ such that there will be no poles in the above equation ($\Bar{A} = \Bar{B}-D = 0$), which immediately takes it to the well-known form of the hypergeometric equation:
\begin{equation}
    z(1-z)F_{zz} + [c - (1+a+b)z]F_{z} - abF = 0~,
\end{equation}
where
\begin{equation}
     \alpha = -i\frac{l^{2}\omega}{2r_{H}}  \hspace{10pt}
    \beta = \frac{1+i\sqrt{6\epsilon-(1+2\,q\,m\,l^{2})}}{2}~.
\end{equation}
\begin{equation}
    a = \alpha + \beta + i\sqrt{C} \hspace{10pt}b = \alpha + \beta - i\sqrt{C} \hspace{10pt}c = 1+2\alpha~.
\end{equation}

In the above, we assume $\epsilon>\frac{1}{6}\,+\frac{1}{3}\,q\,m\,l^{2}$. The general solution to the above equation is as follows \cite{10.5555/1098650}:
\begin{multline}
    R(z) = z^{\alpha}(1-z)^{\beta}\left[C_{1}F(a;b;c;z)\right.\\
    \left.+ C_{2}z^{1-c}F(a-c+1; b-c+1; 2-c; z)\right]~,
\end{multline}
where $C_{1}$ and $C_{2}$ can be determined using our boundary conditions. In fact, we can set $C_{2} = 0$, by demanding purely in-going solution close to the event horizon. We have, therefore:
\begin{equation}
    R(z) = z^{\alpha}(1-z)^{\beta}C_{1}F(a;b;c;z)
\end{equation}

For studying the behavior of the radial function at infinity, we use the following transformation\cite{10.5555/1098650},
\begin{multline}
    F(a,b;c;z)= \frac{\Gamma(c)\Gamma(c-a-b)}{\Gamma(c-a)\Gamma(c-b)} F(a,b;a+b-c+1;1-z)\\
    +(1-z)^{c-a-b} \frac{\Gamma(c)\Gamma(a+b-c)}{\Gamma(a)\Gamma(b)}F(c-a,c-b;c-a-b+1;1-z)
\end{multline}
This means, the radial function at infinity ($z \rightarrow 0 $) can be written in the following form,
\begin{equation}
    R_{FF} \simeq C_{1-}\left(\frac{r}{r_{H}}\right)^{-2\beta} + C_{1+}\left(\frac{r}{r_{H}}\right)^{2\beta-2}~,
\end{equation}
where,
\begin{equation}
    C_{1-}=C_{1}\frac{\Gamma(1+2\alpha)\Gamma(1-2\beta)}{\Gamma(1+\alpha-\beta-i\sqrt{C})\Gamma(1+\alpha-\beta+i\sqrt{C})}
\end{equation}
\begin{equation}
    C_{1+}=C_{1}\frac{\Gamma(1+2\alpha)\Gamma(-1+2\beta)}{\Gamma(\alpha+\beta-i\sqrt{C})\Gamma(\alpha+\beta+i\sqrt{C})}
\end{equation}
Here $C_{1-}$ and $C_{1+}$ terms represent ingoing and outgoing waves respectively. Our boundary conditions at infinity requires the whole solution should vanish at the boundary\cite{Daghigh:2008jz,Cardoso:2015fga}. This means $C_{1-}$ and $C_{1+}$ should vanish which gives us two relations for the overtone number $n$,
\begin{equation}
    -n\,=\,1+\alpha-\beta \pm i\sqrt{C}  \hspace{20pt}-n\,=\,\alpha+\beta \pm i\sqrt{C}
\end{equation}
These relations together with values of above mentioned constants help us to write the quasinormal frequencies as follows:
\begin{equation}
    \omega_{n} = \left|\frac{m}{l} - \frac{\sqrt{M}}{l}\sqrt{6\epsilon-(1+2\,q\,m\,l^{2})}\right| - \frac{2\sqrt{M}}{l}\left(n+\frac{1}{2}\right)i~.
\end{equation}

The real part of the quasinormal modes is modified by anyon correction factors while the imaginary part remains unchanged and negative. This means that the anyon corrections do not have any effects on the stability of black holes.

\section{Experimental setup- Quasinormal modes of anyons from acoustic black holes}
\label{experimentalset-up}

In this section, we explore the feasibility of empirically validating our findings through analog models of gravity that are realizable in a variety of physical systems, including liquid helium and other superfluids. These models exhibit an intriguing parallel with general relativity, where the equation of motion for the system mimics that of a quantum field evolving in an effective curved spacetime background, potentially featuring an acoustic horizon. Within this horizon, quanta of the field, often referred to as phonons, are unable to escape, invoking an intriguing similarity with event horizons in black hole physics.

In particular, for a two-dimensional photon superfluid system, it has been demonstrated that their dynamics can be described by the equation of a massless scalar field within the framework of an acoustic metric \cite{article40}. Furthermore, recent research \cite{PhysRevD.104.104011} has shown that this superfluid system's equation of motion, enriched with a series of corrections that are feasible to implement in laboratory settings, governs the behavior of anyons when immersed in the acoustic metric environment.

This compelling correlation opens the door to the possibility of experimentally investigating the Quasinormal Modes (QNM) of anyons within an analogous setup. It is noteworthy that previous experiments have successfully measured the QNM oscillations of the free surface in a hydrodynamical vortex flow, further validating the feasibility of these analog experiments \cite{PhysRevLett.125.011301}.

Having successfully derived the analytical expression for QNM frequencies of anyons from a BTZ black hole, our research now extends its reach to analogous black hole models with the primary objective of observing these QNM within a laboratory context. To achieve this, we employ a semi-analytic approach based on the Wentzel-Kramers-Brillouin (WKB) approximation \cite{Schutz:1985km}. It is important to note that the potential derived in this setup closely resembles the Regge-Wheeler potential \cite{Regge:1957td}, with the added intricacies arising from the presence of anyons. Additionally, we explore the possibility of employing an analytical approach to obtain an exact spectrum, as detailed in Section \ref{QNMscalar}.

The metric for an acoustic black hole in analogue models of gravity is defined as follows, as described in \cite{article40}:
\begin{multline}
    ds^{2} =\left( \frac{\rho_{0}}{c_{s}} \right)^{2} \left[ -  \left(1 - \frac{v^{2}}{c_{s}^{2}}\right)  \left(c_{s}dt\right)^{2} - 2\frac{v_{r}}{c_{s}} \left(c_{s}dt\right) dr \right.\\
    \left. - 2 \frac{v_{\theta}}{c_{s}} \left(c_{s}dt\right) \left(r\,d\theta\right) +  dr^{2} + \left(r d\theta\right)^{2}\right]~.
    \label{acmetric}
\end{multline}
The metric can be simplified by performing a series of transformations. To begin, we rescale both the space and time coordinates as follows:
\begin{eqnarray}
    r \longrightarrow \frac{\rho_{0}}{c_{s}} \,r \qquad d\,r \longrightarrow \frac{\rho_{0}}{c_{s}}\, d\,r ~,
\end{eqnarray}
\begin{eqnarray}
    t \longrightarrow \rho_{0} \,t \qquad d\,t \longrightarrow \rho_{0}\, d\,t ~.
\end{eqnarray}
%These transformations, together with the assumption $v_{\theta} \sim 0$,  transforms the metric (\ref{acmetric}) into,
These transformations, in conjunction with the assumption that $v_{\theta}$ is approximately zero, result in the transformation of the metric (\ref{acmetric}) into the following form:
\begin{equation}
     ds^{2} = -\left(1 - \frac{v_{r}^{2}}{c_{s}^{2}}\right)  dt^{2} -   \frac{2 v_{r}}{c_{s}}\,dt\, dr +  dr^{2}+r^{2}d\theta^{2}~.
     \label{met5}
\end{equation}
%The above metric can be simplified further by defining $f(r)$,
The metric can be further simplified by introducing the function $f(r)$, which is defined as follows,
\begin{equation}
    f(r)\,=\,1\,-\,\frac{v_{r}^{2}}{c_{s}^{2}} 
    \hspace{10pt}\text{such that,} 
    \hspace{10pt} 
    \sqrt{1\,-\,f(r)}\,=\,\frac{v_{r}}{c_{s}}~,
    \label{fr}
\end{equation}
%in terms of which Eq(\ref{met5}) becomes,
with this definition of $f(r)$, the metric expressed in Eq(\ref{met5}) can be rewritten as:
\begin{equation}
    ds^{2}=-f(r)\,dt^{2}-2\sqrt{1-f(r)}\,dt\,dr+dr^{2}+r^{2}\,d\theta^{2}~.
    \label{metric4}
\end{equation}
Finally, we introduce a coordinate transformation as follows:
\begin{equation}
    dt_{p}=dt+\frac{1}{f}\,\sqrt{1\,-\,f}\,dr~,
\end{equation}
resulting in a metric given by:
\begin{equation}
    ds^{2}=-f(r)\,dt_{p}^{2}+\frac{1}{f(r)}\,dr^{2}+r^{2}\,d\theta^{2}~.
    \label{mmetric}
\end{equation}
%The above metric is also non-rotating and circularly symmetric but differs from the BTZ metric via the functional form of $f(r)$.
This metric, while non-rotating and circularly symmetric, differs from the BTZ metric due to the distinctive form of the function $f(r)$. To determine this functional form, we consider the radial velocity as follows\cite{article40,article22}:
%
%We can find out this functional form by taking radial velocity as follows\cite{article40,article22},
\begin{eqnarray}
    v_{r} = \displaystyle{-\frac{ \, \pi}{k n_{0} \sqrt{r r_{0}}}} ~.
\end{eqnarray}
%Now we have $f(r)$ and the horizon of the acoustic black hole as follows,
We can now express $f(r)$ and the horizon of the acoustic black hole as:
\begin{eqnarray}
     f(r)\,=\,1\,-\,\frac{r_{H}}{r} \qquad r_{H}\,=\,\frac{\pi^{2}}{k^{2}\,n_{0}^{2}\,r_{0}\,c_{s}^{2}}~.
\end{eqnarray}
We derive the equation of motion for anyons from the Lagrangian (\ref{lag1}). To simplify the analysis, we select the vector potential as $A_{\mu}=(0,0,\frac{1}{r})$, as this choice is both simple and general, with $c_2=c_4=0$. The equation of motion can be expressed as:
\begin{equation}
    g^{\mu \nu} \nabla_{\mu}\nabla_{\nu}  \Phi +  2iq g^{\mu \nu}A_{\mu}\partial_{\nu}\Phi - q^2 g^{\mu \nu}  A_{\mu}  A_{\nu}\Phi = 0
    \label{eqan1}
\end{equation}
%To determine the QNM frequencies of anyons in the above background, we use the same ansatz as in Eq.(\ref{ansatz102}). Substituting it in Eq.(\ref{eqan1}), gives the following differential equation for $R$,
To determine the Quasinormal Mode (QNM) frequencies of anyons within this background, we employ the following ansatz,
\begin{equation}
    \Phi\,=\,\frac{R}{\sqrt{r}}\,e^{-i\omega t}e^{i m \theta}
\end{equation}
Substituting this ansatz into Eq.(\ref{eqan1}) results in a differential equation for the radial function $R$:
\begin{multline}
    f^{2}R^{''} + ff^{'}R^{'} + \left(\omega^{2} - \frac{ff^{'}}{2r} + \frac{f^{2}}{4r^{2}} \right. \\
    \left. - \frac{m^2f}{r^{2}} - \frac{2qmf}{r^{3}} - \frac{q^{2}f}{r^{4}} \right)R = 0~.
    \label{eom11}
\end{multline}

The above equation can be simplified further by introducing a new radial coordinate, denoted as $r_{*}$ (the 'tortoise coordinate'), defined as follows:
\begin{equation}
    r_{*}=r+r_{H}\ln{(r-r_{H})}~.
\end{equation}
By substituting the expression of $f(r)$ into Eq.(\ref{eom11}), we obtain the final form of the radial part of the equation, resembling a Schrödinger equation:
\begin{equation}
    \left(\partial^{2}_{r_{*}}  + Q(r)\right)R = 0~,
\end{equation}
%where,
where the function $Q(r)$ is defined as:
\begin{equation}
    Q(r) \equiv \omega^{2}-V(r)~,
    \label{QNMA}
\end{equation}
and the potential $V(r)$ is given by:
\begin{align}
    V(r) = f(r)\,\left(\frac{4m^{2}-1}{4 r^{2}} + \frac{3 r_{H}}{4 r^{3}}\right)+\frac{2\,q\,m\,f(r)}{r^{3}}+\frac{q^{2}f(r)}{r^{4}}~.
\end{align}
It can be demonstrated that $-Q(r)$ attains its maximum value at $r = r_{0}$. Moreover, by employing the Wentzel–Kramers–Brillouin (WKB) approximation, it is possible to establish a relationship between $Q$ and its second derivative at $r_0$, denoted as $Q(r_0)$ and $Q''(r_0)$, respectively \cite{Schutz:1985km}:
\begin{equation}    
Q_{0} = i\sqrt{2Q^{''}_{0}}\left(n + \frac{1}{2}\right)~.
\label{wkb}
\end{equation}
Subsequently, the Quasinormal Modes (QNM) can be extracted from Eq.(\ref{QNMA}) as follows:
\begin{eqnarray}
    \omega_{n}=\left\{V(r_{0})+i(2Q_{0}^{''})^{\frac{1}{2}}\left(n+\frac{1}{2}\right)\right\}^{1/2}~.
    \label{QNMB}
\end{eqnarray}
Equation (\ref{QNMB}) characterizes the QNM of anyons within an analogue black hole, and these QNMs hold the potential for experimental verification in a laboratory setting. Although it is possible to express the above equation in terms of $r_H$ and $m$, such expressions tend to be complex and do not offer additional valuable insights.

\section{Conclusion}
\label{conclusionsection}

In this work, we have undertaken a comprehensive investigation of Quasinormal Modes (QNM) in the context of anyons, particles with intermediate statistics between a boson and a fermion, 
in the realm of $(2+1)$-dimensional BTZ black holes. 
We have successfully derived an expression for these QNM.
Our findings reveal a distinct characteristic of these QNM: while the imaginary part remains unaltered in comparison to the QNM of a massless scalar field in similar backgrounds, the real part is notably modified. This observation underscores the unique interplay between anyons and gravitational systems, shedding light on the intriguing dynamics that underlie these phenomena.

We have taken a step further by exploring the practicality of experimentally observing these QNM in analogue black holes, particularly focusing on acoustic black holes. By applying a semi-analytic approach, the Wentzel-Kramers-Brillouin (WKB) method, we have derived the QNM frequencies for anyons in the context of acoustic black holes. It is crucial to acknowledge that the WKB approach, as an approximation method, opens avenues for future refinement and accuracy enhancement through the consideration of higher-order terms. This extension is pivotal as it paves the way for even more precise predictions and interpretations of the experimental outcomes.

The implications of this research extend beyond theoretical exploration and serve as a foundation for experimental validation. The experimental verification of our current results can be viewed as tangible evidence not only for the existence of anyons but also for the existence of their QNM within spacetimes featuring horizons. This empirical confirmation would offer a critical link between the abstract theoretical constructs and real-world phenomena, providing a deeper understanding of the enigmatic nature of anyons in black hole environments. As we continue to push the boundaries of our knowledge in this domain, it will be instrumental to extend our investigations to other analogue spacetimes characterized by additional attributes such as angular momentum or charge. This expansion of our research scope promises to unveil new facets of the interactions between anyons and horizons, enriching our comprehension of these intriguing quantum systems within the framework of general relativity.

\section{Acknowledgement}
This work was supported by the Natural Sciences and
Engineering Research Council of Canada. 
V.C was supported by research fellowships from the Council of Scientific and Industrial Research, India.\\
\textit{Data availability} : No data associated in the manuscript.
\nocite{churilova2020quasinormal}\nocite{PhysRevD.101.124012}\nocite{2004}\nocite{doi:10.1126/science.aaz5601}\nocite{article32}\nocite{article40}\nocite{article22}\nocite{2017}\nocite{2008qnm}\nocite{PhysRevD.93.044048}\nocite{Schutz:1985km}\nocite{Panotopoulos:2018can}\nocite{PhysRevD.105.044015}\nocite{PhysRevLett.123.111102}\nocite{PhysRevLett.69.1849}\nocite{Sen:1993qc}\nocite{Rao1992AnAP}\nocite{Chandrasekhar:1975zza}\nocite{Regge:1957td}\nocite{PhysRevD.63.124015}\nocite{https://doi.org/10.48550/arxiv.hep-th/9508151}\nocite{Mathur:2005zp}\nocite{Almheiri:2013hfa}\nocite{doi:10.1142/S0218271821500693}\nocite{PhysRevD.104.104011}\nocite{PhysRevD.64.064024}\nocite{PhysRevLett.125.011301}\nocite{AGHAEIABCHOUYEH2018240}\nocite{Luo:2017ksc}\nocite{RevModPhys.80.1083}\nocite{Field_2018}\nocite{Nakamura_2020}\nocite{PhysRevD.97.084034}\nocite{https://doi.org/10.48550/arxiv.2210.14973}\nocite{PhysRevD.1.2870}\nocite{PhysRevD.63.124015}\nocite{10.5555/1098650}\nocite{Daghigh:2008jz}\nocite{Cardoso:2015fga}
\bibliographystyle{ieeetr}
\bibliography{main}
\end{document}